\documentclass[aps,prl,preprint,showpacs,groupedaddress]{revtex4}

\pdfoutput=1
\usepackage{amsmath}
\usepackage{amssymb}
\usepackage{graphics}
\usepackage{wrapfig}
\usepackage[dvips]{graphicx}
\graphicspath{{./},{./pictures/}}
\usepackage{color}

\begin{document}

\title{Gravitomagnetic relativistic effects on turbulence }

\author{Demosthenes Kivotides
}

\affiliation {University of Strathclyde Glasgow
}

\date{\today}

\begin{abstract}
The dynamics of fluid-matter under the influence of gravitomagnetic fields
are formulated and solved for the case of fully developed turbulence.
Gravitomagnetic effects reduce the vortical complexity and nonlinearity of turbulence, even leading to its extinction
within large volumes, and generate departures from Kolmogorov turbulence scalings, that are explained via a combination of
dimensional and exact analysis arguments.
\end{abstract}


\maketitle

\section{Introduction}

Understanding strongly nonequilibrium dynamics and emergence in 
basic field theories is a main task of statistical and nonlinear physics.  
In {\it electrodynamics}, an efficient way to tackle far-out of equilibrium processes
is to focus on the macroscopic scales, and, by taking the {\it magnetic} and {\it small-velocity} limits of the theory,
to formulate the magnetohydrodynamic (MHD) equations of motion. Strongly nonequilibrium states in these equations, 
known as MHD turbulence, are well understood, with many, fully-resolved numerical calculations having become available \cite{biskamp,kivot_plas2}. 
At small enough temperatures for the formation of quantum mechanical, atomic bound states \cite{pert,efimov,salzmann}, we obtain {\it low-energy electrodynamics} \cite{stone}, and
MHD reduces to standard hydrodynamics and turbulence \cite{davidson}.
There are similar approaches for Newtonian {\it gravitodynamics} \cite{galactic}, but the problem of relativistic 
gravitational turbulence is not equally well developed.\\

Indeed, on one hand, turbulence is strongly nonlinear, and evolves over a continuum range of scales,
which need to be well resolved to correctly capture its statistical structure. A key requirement is the resolution
of the, all important, dissipation processes, whose efficiency peaks at high wavenumbers of the energy spectrum, where also most of the strain resides.
On the other hand, the {\em fully relativistic} problem is so computationally complex, that calculations satisfying these standards
are very difficult to perform. This research aims to formulate a hydrodynamic model of self-gravitating matter,
which although {\em not fully relativistic}, retains some (conceptually important for gravitational theory) relativistic effects, known as gravitomagnetism.
In this way, model generality is traded for computational quality.
Another aim is to perform actual calculations with the simplified model, indicate particular gravitational effects on turbulence structure,
and explain the resulted statistical phenomenology via scaling arguments.\\

Gravitomagnetism is far less well studied than ``gravitoelectricity" (Newtonian gravity) \cite{wheeler}.
Indeed, in Galilean invariant theory of gravity, a rotating, massive sphere produces identical gravitational fields with a stationary one. But in relativity, and in the weak field, slow motion limit,
the rotation of a massive sphere adds to the standard Schwarzschild field of a stationary sphere a gravitomagnetic element.
A chief motive for this enquiry, is the fact that turbulence, when seen as a strongly out of equilibrium, nonlinear system, is dominated 
by linear vortical structures (quasi-defects) \cite{davidson}, hence, rotating matter is the {\em very essence} of its physics.
In other words, turbulence is the arena of the most complicated gravitomagnetic phenomena, which are the focus of this study.\\

Apart from their implications for gravity and turbulence theories, the results could, possibly, also benefit astrophysical investigations.
Indeed, in {\em neutron stars}, gravitomagnetic effects were shown to affect precession rates by about $10\%$ \cite{levin},
and there is need to improve the quality of employed phenomenological {\em turbulence models} in {\em relativistic hydrodynamics}
investigations of their differential rotation and mergers \cite{shibata,radice}. Moreover, some of the insight into the structure
of gravitomagnetic turbulence provided here, could inform astrophysical investigations of 
interstellar and inter-galactic medium turbulence, where it is important to understand turbulence vorticity dynamics \cite{ryu,schuecker},
and of large scale flow in galactic supercluster assemblies,
where there is need for inclusion of nonlinear, self-consistent gravitodynamic effects \cite{kremer,laniakea,somak,kitaura}.\\

\section{Self-gravitating particle systems at large-scale, weak-field and slow-motion limits}

We aim to understand the large-scale dynamics of a microscopic theory
describing the self-consistent interactions between gravity and a discrete system (dust) of $N$ {\it spinless} point particles (which,
depending on the application setting, could be interstellar or inter-galactic medium particles, stars or galaxies) \cite{thorne,zee,wheeler}
\begin{equation}\label{micro}
\begin{split}
R^{\mu\nu}-\frac{1}{2} g^{\mu\nu} R=\frac{8 \pi G}{c^4} T^{\mu\nu},   \nonumber \\
\frac{{\rm d}u_p^{\mu}}{{\rm d}\tau}+\Gamma_{\nu \lambda}^{\mu} u_p^{\nu} u_p^{\lambda}=0,  \nonumber \\
T^{\mu \nu}(x^{\lambda})=\sum_{p=1}^{N} m c \int{\rm d}\tau \frac{\delta^{(4)}[x^{\lambda}-r_p^{\lambda}(\tau)]}{\sqrt{-g}} u_p^{\mu} u_p^{\nu}, \nonumber
\end{split}
\end{equation}
\noindent where $\sum$ indicates summation over $p=1,...,N$ particles in the system (Einstein summation convention
not valid for $p$), $\mu, \nu, \lambda=1,2,3,4$,
$m$ is the particle mass,
$u_p^{\mu}={\rm d}r_p^{\mu}/{\rm d}\tau$ is the particle four-velocity,
${\rm d}\tau=\sqrt{-ds^2}/c$ is the proper time along particle trajectories, 
$ds^2=g_{\mu \nu} {\rm d}x^{\mu} {\rm d}x^{\nu}$ is the differential spacetime interval,
$g_{\mu \nu}$ is the metric tensor, $g$ is the determinant of $g_{\mu \nu}$,
$R^{\mu\nu}$ is the Ricci curvature tensor, $R$ is the scalar curvature, $T^{\mu \nu}$ is the energy-momentum tensor,
and $\Gamma_{\nu \lambda}^{\mu}$ are the Christoffel symbols. $G$ is the gravitational constant and $c$ is the 
speed of light. The first equation is the Einstein field equation that results from gauging the symmetry with 
respect to displacement tranformations, and, since the particles are spinless, is a 
complete description of gravitational interactions in the system \cite{feynman,blago,rando}.
The second equation indicates that gravitational effects are globally equivalent to inertia in curved spacetimes
or, locally, to the effect of a non-inertial reference frame in flat, special-relativistic spacetime,
a feature that is going to play an important role in the derivation of the hydrodynamic model.\\

For large $N$ typical of applications, it is practically 
difficult to calculate the evolution of the microscopic system over spacetime scales large
enough to capture the phenomenology of fully developed turbulence. Hence, we coarse-grain the dynamics
to obtain first the Einstein-Boltzmann system \cite{kinetic1,kinetic2,kinetic3}, and, further on, the 
Einstein-Navier-Stokes System (ENSS) of {\em relativistic} compressible fluid dynamics \cite{ferrarese,disconzi,geroch,semelin,rezzolla},
where the Einstein equation is now the ensemble average of the corresponding microscopic equation. 
Notably, since gravitational systems are long-range interacting systems, the Boltzmann equation is not meant here
in the sense of its familiar version for dilute gases, but in the general statistical mechanical sense
of a kinetic equation with a collisional term. Then, the hydrodynamic approximation is valid at large times (in units of
a relevant relaxation timescale). The latter depends on the particular system and modeling purposes \cite{longrange}.
It could be the fast timescale $\tau_v$ of {\em violent} relaxation, which is independent of the number of particles $N$, 
or the longer {\em collisional} relaxation timescale $\tau_c=N^{\delta}$ (where ${\delta}$
is a system-dependent exponent).
A typical $\tau_c$ example is the Chandrasekhar relaxation timescale in stellar systems.  
{\em Local equilibria} based on  $\tau_v$ are solutions of the Vlasov equation, and 
those based on $\tau_c$ of the Boltzmann equation \cite{kinetic1,longrange}.
Navier-Stokes type of diffusion has been applied to cosmological \cite{gurbatov}, and interstellar medium dynamics \cite{semelin}.
Although the ENSS system incorporates the physics of fully developed turbulence,
the computational complexity is so high, that fully resolved calculations are not presently available. 
So, in this work, we proceed with a theoretical formulation in between Newton and Einstein gravities
which retains significant relativistic effects, whilst allowing {\it routine}, fully resolved turbulence calculations.\\

In this context, we take the {\it weak field}, {\it slow motion} limit of ENSS, which reduces the Einstein equation
to a type of gravitational Maxwell theory \cite{poisson}, and the relativistic compressible Navier-Stokes equation to its Newtonian counterpart.
The latter have already been employed in {\em conjunction} with gravitational Maxwell theory
in the study of accretion disks around massive astronomical objects (\cite{wheeler}, page 327). A thorough discussion
of the Newtonian Navier-Stokes limit is available in \cite{rezzolla} (page 294). Hence, a simplified model is
{\begin{center}
\begin{eqnarray}\label{full}
\partial_i E_i^g=-4 \pi G \rho, \nonumber \\
\epsilon_{ijk} \partial_j E_k^g=-\partial_t B_i^g, \nonumber \\
\partial_i B_i^g=0, \nonumber \\
\epsilon_{ijk} \partial_j B_k^g=-\frac{16 \pi G}{c^2} \rho u_i+\frac{4}{c^2} \partial_t E_i^g, \nonumber \\
\partial_t \rho+\partial_i (\rho u_i)=0, \nonumber \\
	\partial_t (\rho u_i)+ \partial_j (\rho u_i u_j)=\rho (E_i^g+\epsilon_{ijk} u_j B_k^g)-\partial_i p + \partial_j \sigma_{ij},  \nonumber 
\end{eqnarray} 
\end{center}}
where the last two equations are the compressible Navier-Stokes (NS) equations,
$i, j, k=1,2,3$, $\epsilon_{ijk}$ is the Levi-Civita symbol, $E^g$ is the gravitoelectric field (i.e., the Newtonian gravitational
field), $B^g$ is the gravitomagnetic
field, $u$ is the fluid velocity, $\rho$ is the density of fluid mass-energy, $p$ is the scalar pressure,
and $\sigma_{ij}$ is the viscous stress tensor. The corresponding scalar and vector potentials for $E^g$ and $B^g$ are the  
$g_{00}$ and $g_{0i}$ ($i=1,2,3$) $g_{\mu\nu}$ components. The remaining six components $g_{ij}$ become irrelevant,
since they generally describe the geometries of curved three dimensional spaces (slices of constant $ct$), which,
for {\it weak fields} and {\it low velocities} of interest here, become flat ($R^{ij} \equiv 0$).
Hence, the tensor-field theory of gravity is reduced to a vector one, but the causal structure of Minkowski spacetime
(hence, also relativistic effects) are preserved in the guise of gravitomagnetism.\\

\section{Gravitomagnetic fluid dynamics equations and their scaling}

In the context of our investigation of self-gravitating, {\em homogeneous, isotropic} turbulence, the above system will be further simplified.
Indeed, it is known that compressibility effects cause deviations from Kolmogorov scalings of incompressible turbulence
when turbulent eddies are in transonic or supersonic motion relative to each other \cite{shear}. The dimensionless number quantifying this notion
is the Mach number of turbulence $M_t=u^{\prime}/c_s$, where $u^{\prime}$ is the intensity of turbulent velocity fluctuations, and 
$c_s$ is the medium's speed of sound \cite{shear}. In other words, $M_t$ measures the ratio between turbulent kinetic energy and thermal energy, and the latter
ought to be a significant proportion of the former, for compressibility effects to become important. There is observational evidence that many astrophysical flows
could be treated as incompressible. For example, analysis of turbulent gas pressure maps in the intra-cluster medium revealed that 
pressure fluctuations are consistent with {\em incompressible, Kolmogorov} turbulence \cite{ryu,schuecker}. This follows
directly from the Mach number criterion, since, for the Coma cluster \cite{ryu,schuecker}, it is estimated that
$\epsilon_{turb} \geq 0.1 \epsilon_{th}$,
where $\epsilon_{turb}$ is the kinetic energy density of turbulence, and $\epsilon_{th}$ is the fluid's thermal energy density,
hence, the turbulent fluctuations are subsonic. These suggest that it would be useful to take, in the full model, the incompressible limit
of the Navier-Stokes equation. That this is a meaningful limit of a {\em relativistic theory} 
has been demonstrated in many different works \cite{kip,veronika,bhatta,strominger}.
This step allows also an important simplification of the gravity part of the equations: in a {\em homogeneous, constant mass-energy density matter system}, the
gravitoelectric field $E^g$ becomes dynamically irrelevant, hence, the time derivative of $B^g$ becomes zero,  the ``displacement
current" $\frac{4}{c^2} \partial_t E_i^g$ can be dropped from the equation for the {\it curl} of $B^g$, and 
$E^g$ can be dropped from the NS equation.
Notably, gravitodynamics is more nonlinear than electrodynamics, since, in the latter, electric current $J^e$
includes the electric charge density $\rho^e$ which differs from the fluid density $\rho$, whilst, in the
former, gravitational current $J^g$ is identical to the {\it fluid momentum}. This identification of
gravitational charge with inertial mass, or, as indicated above, of gravitational effects with inertia,
allows the two $B^g$ equations to form, together
with the incompressible NS, a {\it closed} system of differential equations that can be autonomously solved,
{\begin{center}
\begin{eqnarray}\label{lim}
\partial_i B_i^g=0, \nonumber \\
\epsilon_{ijk} \partial_j B_k^g=-\frac{16 \pi G}{c^2} \rho u_i, \nonumber \\
\partial_i u_i=0, \nonumber \\
\partial_t u_i=\epsilon_{ijk} u_j B_k^g-\partial_i\bigg(\frac{p}{\rho}+\frac{u_j u_j}{2}\bigg)+\epsilon_{ijk} u_j \omega_k+\nu \partial_j \partial_j u_i.  \nonumber 
\end{eqnarray}
\end{center}}
\noindent It is important to note, that since we aim to study homogeneous, isotropic turbulence here, the above system refers exclusively to {\em fluctuating quantities}.
Although the equations look similar to analogous equations in the electrodynamic magnetic limit \cite{bellac,bagchi}, their physical motivation is {\em very different}.
In electrodynamics, there are negative and positive charges, hence, it is possible to realize the magnetic limit, by having approximately zero charge densities, but significant
currents, which are responsible for neutralizing the charges \cite{qgp}. In gravity, on the other hand, there cannot be zero mass-energy densities, but
in {\em homogeneous} incompressible  media, and in the weak-field, slow-motion limit, gravitomagnetism is, as explained above, all there is, and only gravitational effects 
induced by the flow of matter are observable.
Certainly, gravitoelectric effects are very important in {\em inhomogeneous} incompressible systems, as are, for example, variable density (stratified) media, 
or accretion disks around massive central objects \cite{wheeler} (page 327).\\ 

\noindent To obtain a scaled system of equations, we define the constant $\beta\equiv {16 \pi G}/{c^2}$,
and use it to scale $B^g$ as $\widetilde{B}^g\equiv B^g/\sqrt{\beta \rho}$, and define 
the scaled gravitational current $\widetilde{J}_i^g=\epsilon_{ijk} \partial_j \widetilde{B}_k^g$. 
Notably, $\sqrt{\beta \rho}$ has units of ${\rm cm}^{-1}$,
$\widetilde{B^g}$ has units of velocity ${\rm cm}\ {\rm s}^{-1}$, and $\widetilde{J}^g$ has the units ${\rm s}^{-1}$ of flow vorticity
$\omega_i\equiv \epsilon_{ijk} \partial_j u_k$.
Finally, by taking the {\it curl} of $\epsilon_{ijk} \partial_j B_k^g$, we
arrive at the {\em scaled} gravito-magneto-hydrodynamic (GMHD) equations
\begin{eqnarray}\label{gmhd}
\partial_j \partial_j \widetilde{B}_i^g=\sqrt{\beta \rho} \omega_i, \nonumber \\
\partial_i u_i=0, \nonumber \\
\partial_t u_i=-\partial_i\bigg(\frac{p}{\rho}+\frac{u_j u_j}{2}\bigg)+\epsilon_{ijk} u_j \omega_k-\epsilon_{ijk} \widetilde{B}^g_j \widetilde{J}^g_k+\nu \partial_j \partial_j u_i,  \nonumber 
\end{eqnarray}
where it is instructive to compare the equation for $\widetilde{B}_i^g$ with the equation for the velocity vector potential $\psi$, $\partial_j \partial_j \psi_i=-\omega_i$, where
the difference in signs is due to the negative sign in the left hand side of the equation for the {\it curl} of $B^g$. Notably, Lamb force
$\epsilon_{ijk} u_j \omega_k$ is the vector product of velocity with its vorticity, hence, since the gravitational current
is the momentum, the novelty of gravitational effects (in comparison with Lamb force effects) depends on the relative orientation of $\omega$ and $B^g$ vectors.
In this form of the NS equation, $\epsilon_{ijk} \widetilde{B}^g_j \widetilde{J}^g_k$ term encodes genuine relativistic gravitational effects
on flow structures. Indeed, it is straightforward to demonstrate \cite{franklin} (page 89), that by combining Newtonian gravity with Lorentz transformations,
and demanding identical physical predictions for different inertial frames, one ``discovers" gravitomagnetism. In other words, the latter is a direct consequence of the relativity principle. 
It is helpful to write the vorticity dynamics equation
\begin{eqnarray}\label{vorticity}
\partial_t \omega_i+u_j \partial_j \omega_i-\omega_j \partial_j u_i-\widetilde{B}^g_j \partial_j \widetilde{J}^g_i+\widetilde{J}^g_j 
\partial_j \widetilde{B}^g_i-\nu \partial_j \partial_j \omega_i=0,  \nonumber 
\end{eqnarray} 
where the sum of the four inner terms could be succinctly written in terms of Lie derivatives as ${\mathcal L}_u \omega-{\mathcal L}_{\widetilde{B}^g} \widetilde{J}^g$.
It is straightforward then to define a gravitational interaction parameter $N^g$, that measures gravitomagnetic effects in units of fluid inertia,
$N^g=\vert {\mathcal L}_{\widetilde{B}^g} \widetilde{J}^g \vert/\vert {\mathcal L}_u \omega \vert$. By inserting typical values of the various quantities
in this expression, we obtain $N^g= \ell_{g} \ell_{h} \beta \rho$, where $\ell_g$ and $\ell_h$ are length scales typical of gravitational and hydrodynamic-variable {\it gradients}
correspondingly. 
The other important parameter, that measures {\it nonlinear}, inertial, nonequilibrium processes in units of linear, viscous, nonequilibrium processes,
is the Reynolds number, $Re=u^{\prime} l/\nu$, where  $u^{\prime}=\sqrt{\langle u^2 \rangle}$ is the turbulent intensity of $u$,
and $l$ is the {\it integral length scale} which measures the size of turbulence-energy containing eddies \cite{davidson}.\\

\section{Numerical methods and finite precision arithmetic}

The model is solved with a staggered grid, fractional step, projection, finite volume, numerical method \cite{projection, kivot_plas1}.
Spatial partial derivatives are computed with second
order accurate schemes. 
An implicit, second order accurate in time, Crank-Nicolson (CN) scheme
is applied to the viscous/diffusion terms, whilst all other terms evolve via
an explicit, third order accurate in time, low storage Runge-Kutta (RK) method.
The CN scheme is incorporated into the
RK steps and the method becomes a hybrid RK/CN scheme.
Flow incompressibility is enforced by projecting the velocity onto the space of
divergence-free vector fields (Hodge projection).
The time-steps are adaptive, limited by the Courant-Friedrichs-Lewy condition,
and resolve the viscous processes in the flow. The gravitomagnetic field is computed
self-consistently from the corresponding Poisson equation with Fast Fourier Transform
methods. The algorithmic approximation of this numerical analysis
adds {\it finite-precision} arithmetic round-off error to analytic truncation error: within the employed floating point number set $\mathbb{F}$,
the distance between $1$ and the next larger floating point number is $\epsilon_m=0.222 \times 10^{-15}$.
The smallest and largest numbers that can be represented are
$2.2 \times 10^{-308}$ and $1.8 \times 10^{308}$ correspondingly.
The algorithm arithmetic employs the {\it round to nearest even} rounding mode \cite{validated,kornerup}.
The computational domain is a cube discretized into $256^3$ grid cells. In all calculations, the dissipation scales
are fully resolved.\\

\section{Homogeneous, isotropic, gravitomagnetic turbulence}

First, we set up a steady-state, homogeneous, isotropic, pure NS turbulence with Taylor Reynolds number
$Re_{\lambda}=u^{\prime} \lambda/\nu\approx 80$, and then we switch on gravity. 
Here, $\lambda$ is the Taylor microscale $\lambda^2=15 \nu \langle u_i u_i \rangle/3 \epsilon$,
where $\epsilon= 2 \nu \langle S_{ij} S_{ij} \rangle$ is the
rate of turbulence energy dissipation, and
$S_{ij}=\frac{1}{2}(\partial_j u_i+\partial_i u_j)$ is the {\it strain rate tensor} \cite{davidson}.
To achieve steady state, viscous dissipation action is compensated by Lundgren's linear forcing \cite{carroll,kivot_pof}.
$\lambda$ is a good representation of the length scale where most of turbulent strain takes place,
so it is a good candidate for $\ell_h$, since the latter has to characterize vortex stretching. On the other hand, 
the solution of the Poisson equation for $\widetilde{B}^g(x)=-\frac{\sqrt{\beta \rho}}{4 \pi} \int 
\frac{\omega(x^{\prime}) {\rm d}x^{\prime}}{\vert x^{\prime}-x \vert}$
indicates that $\widetilde{B}^g$ is formed by the weighted sum of neighbouring-vorticity contributions,
so it could be expected that gravitomagnetic gradients would scale with the largest correlation
length in the system, which in turbulence case is the integral length scale $l$, that measures the size of large eddies.
Hence, $N^g= l \lambda \beta \rho$. This intuition is fully supported by the computational solutions.
Some important time scales are 
the viscous time scale $\tau_d=(\Delta x)^2/(6 \nu)$,
where $\Delta x$ is the computational grid size, the 
gravitational time scale $\tau_g^{N^g}=l/(\widetilde{B}^g)^{\prime}$,
where $(\widetilde{B}^g)^{\prime}=\sqrt{\langle (\widetilde{B}^g)^2 \rangle}$ is the turbulent intensity of $\widetilde{B}^g$,
and the time scale of energy containing motions, $\tau_e^{N^g}=l/u^{\prime}$.\\

\begin{figure*}
\begin{minipage}[t]{0.99\linewidth}
\begin{tabular}[b]{ccc}
\includegraphics[width=0.33\linewidth]{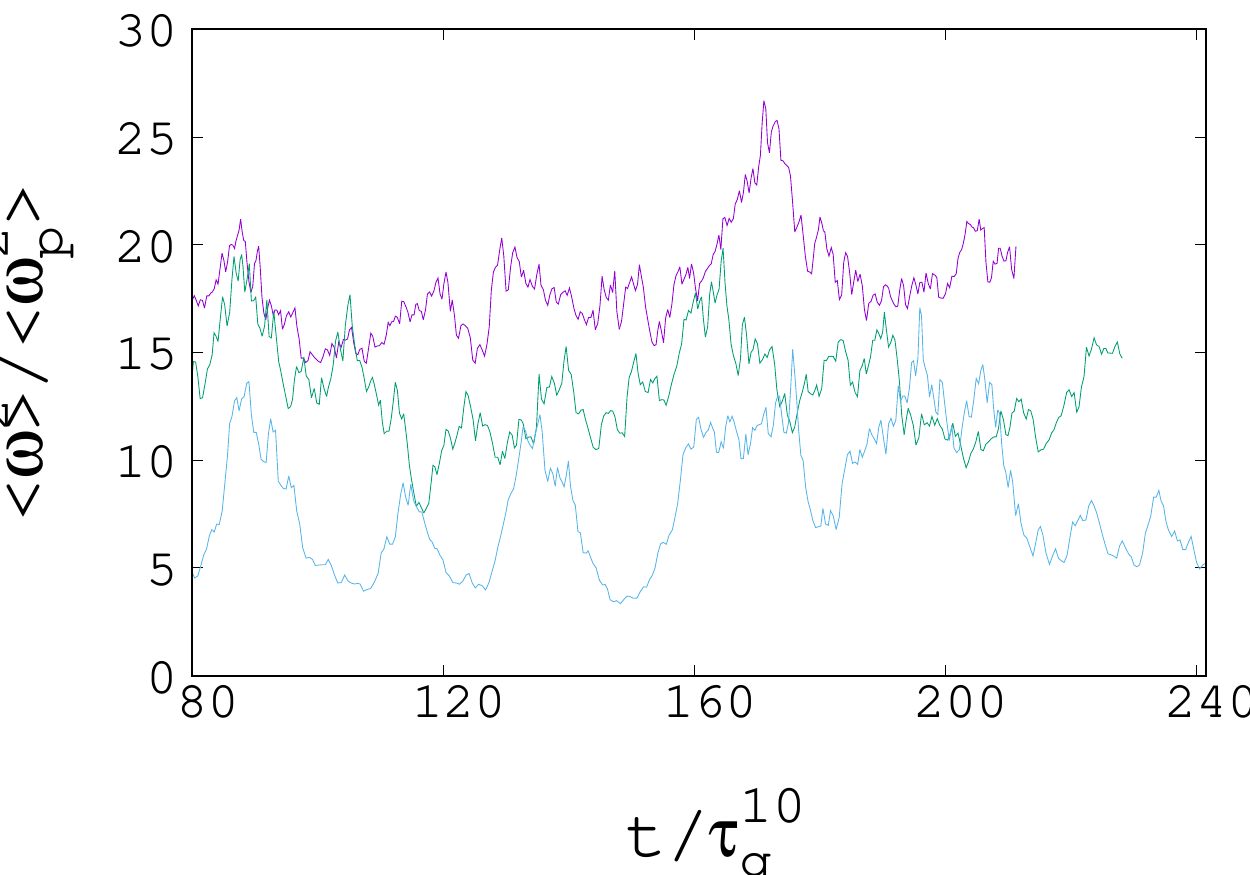}&
\includegraphics[width=0.33\linewidth]{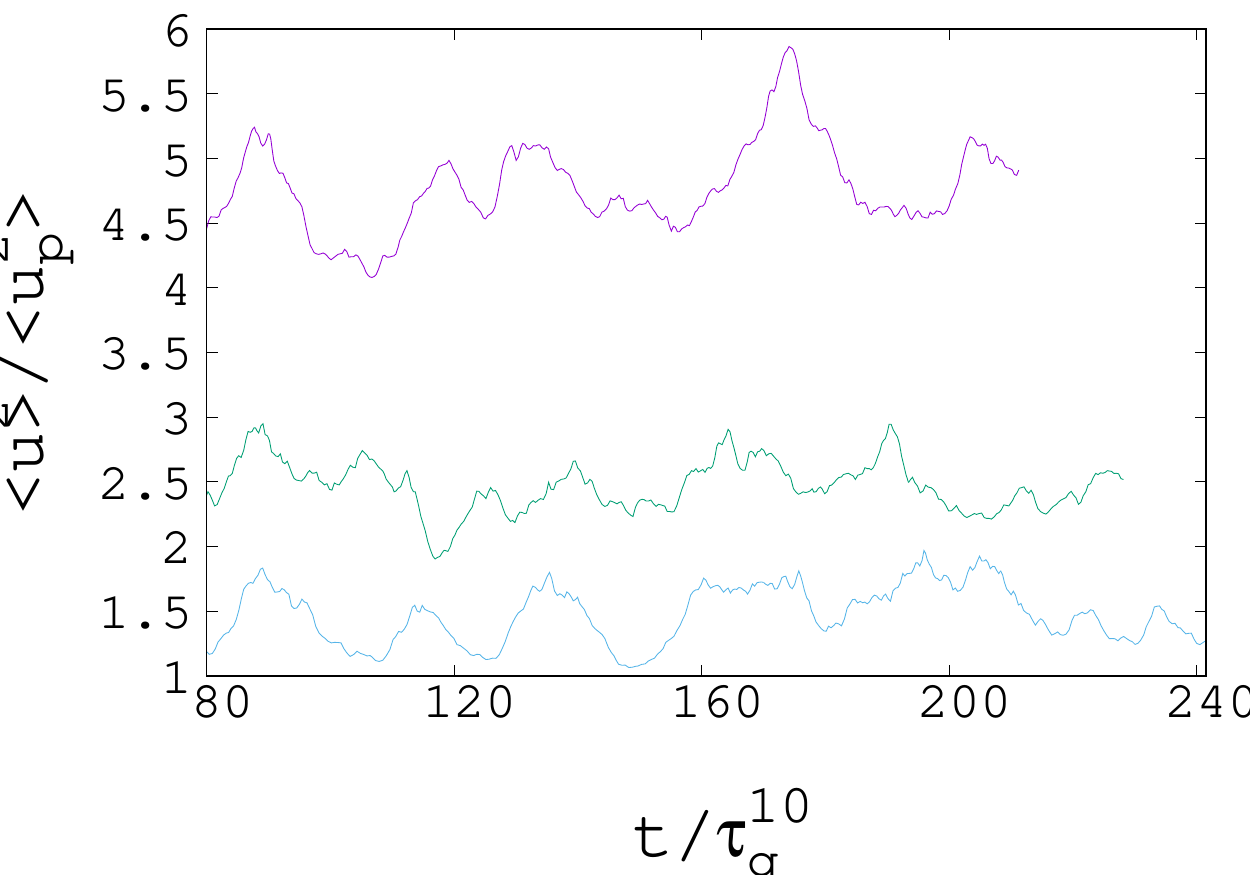}&
\includegraphics[width=0.33\linewidth]{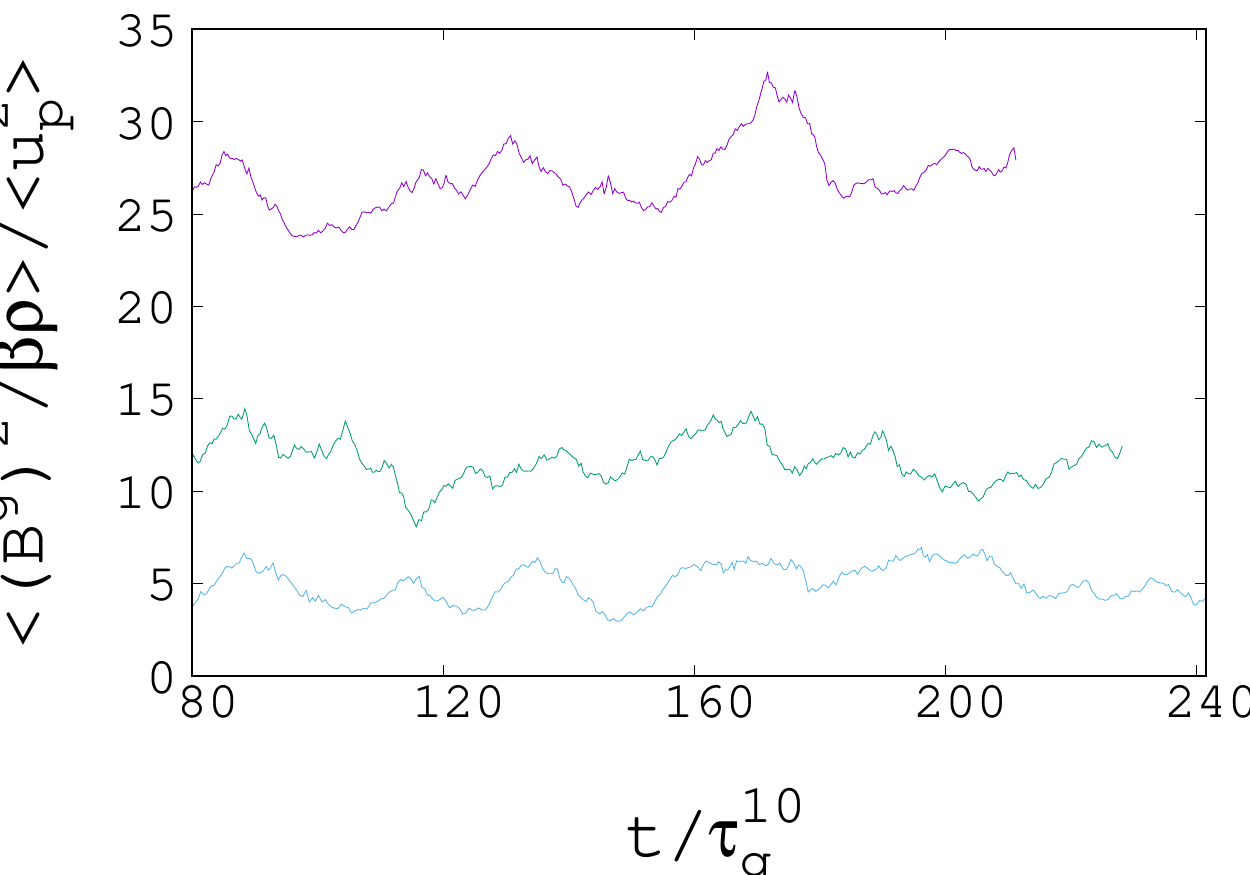}\\
\end{tabular}
\caption{\label{2}
Steady state turbulence averages for $N^g$=[10,20,40]. From left to right: vorticity, velocity and gravitomagnetic field mean squares. 
$N^g$ increases from top to bottom curves. The shown time period corresponds to several dozens of $\tau_e^{N^g}$. The shown quantities
are scaled with the corresponding {\it pure} turbulence values.
}
\end{minipage}
\end{figure*}

To achieve a statistical steady state of turbulent gravity-matter interactions, we continue compensating viscous dissipation
after enabling gravity. 
Starting from $N^g=0.01$ and increasing the strength of gravitational effects, we find important effects
close to $N^g=10$, so we performed three extended-time calculations for $N^g=[10,20,40]$.
Notably, $N^g$ is indicative of the coupling-strength between matter and gravity, which is a dynamical quality not to be
confused with the strength of the resulting gravitational fields.
The relations between the various time scales are $\tau_g^{10}= 9.8 \tau_d$, $\tau_g^{10}= 0.41 \tau_e^{10}$,
$\tau_g^{20}=1.52 \tau_g^{10}$, and $\tau_g^{40}=2.4 \tau_g^{10}$, $\tau_e^{20}=1.37 \tau_e^{10}$, $\tau_e^{40}=1.81 \tau_e^{10}$. 
\begin{figure*}
\begin{minipage}[t]{0.99\linewidth}
\begin{tabular}[b]{cc}
\includegraphics[width=0.49\linewidth]{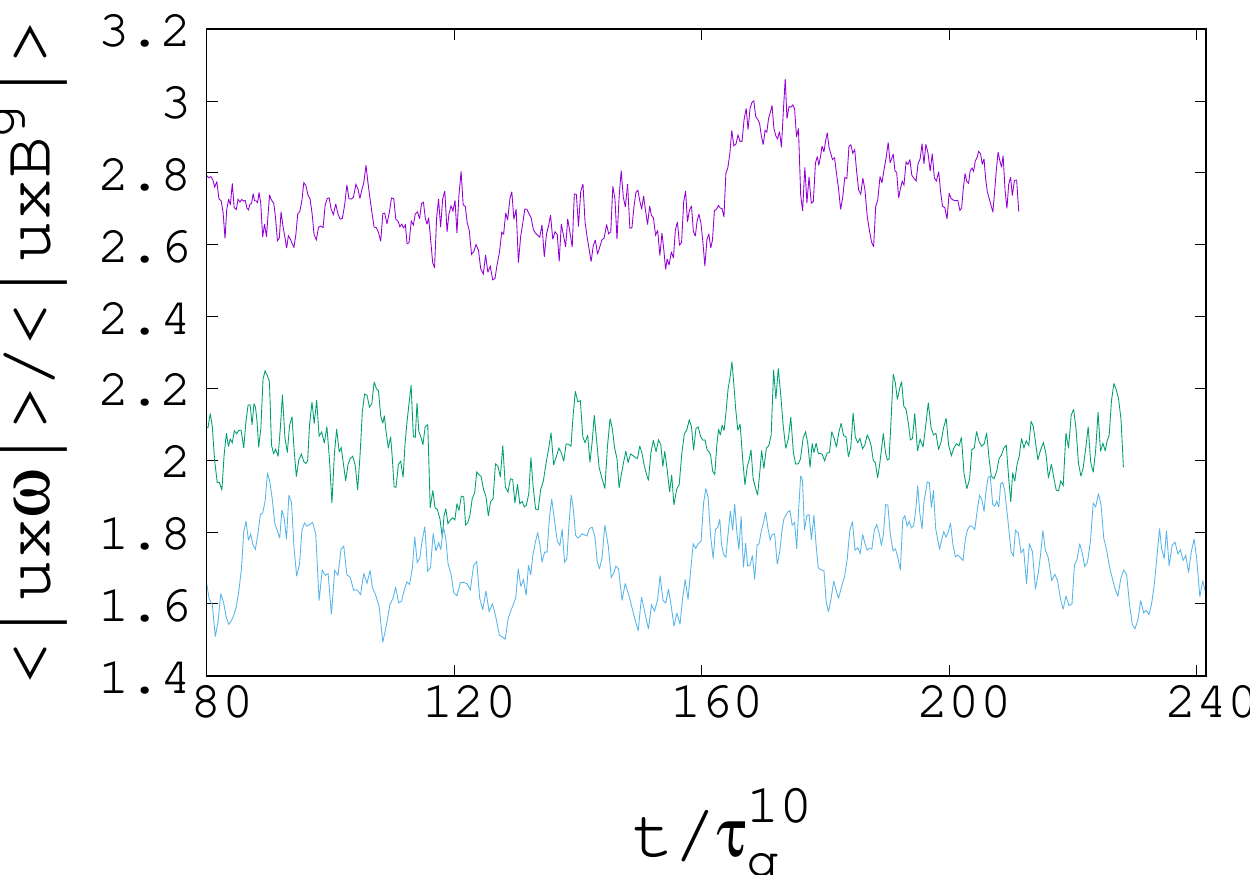}&
\includegraphics[width=0.49\linewidth]{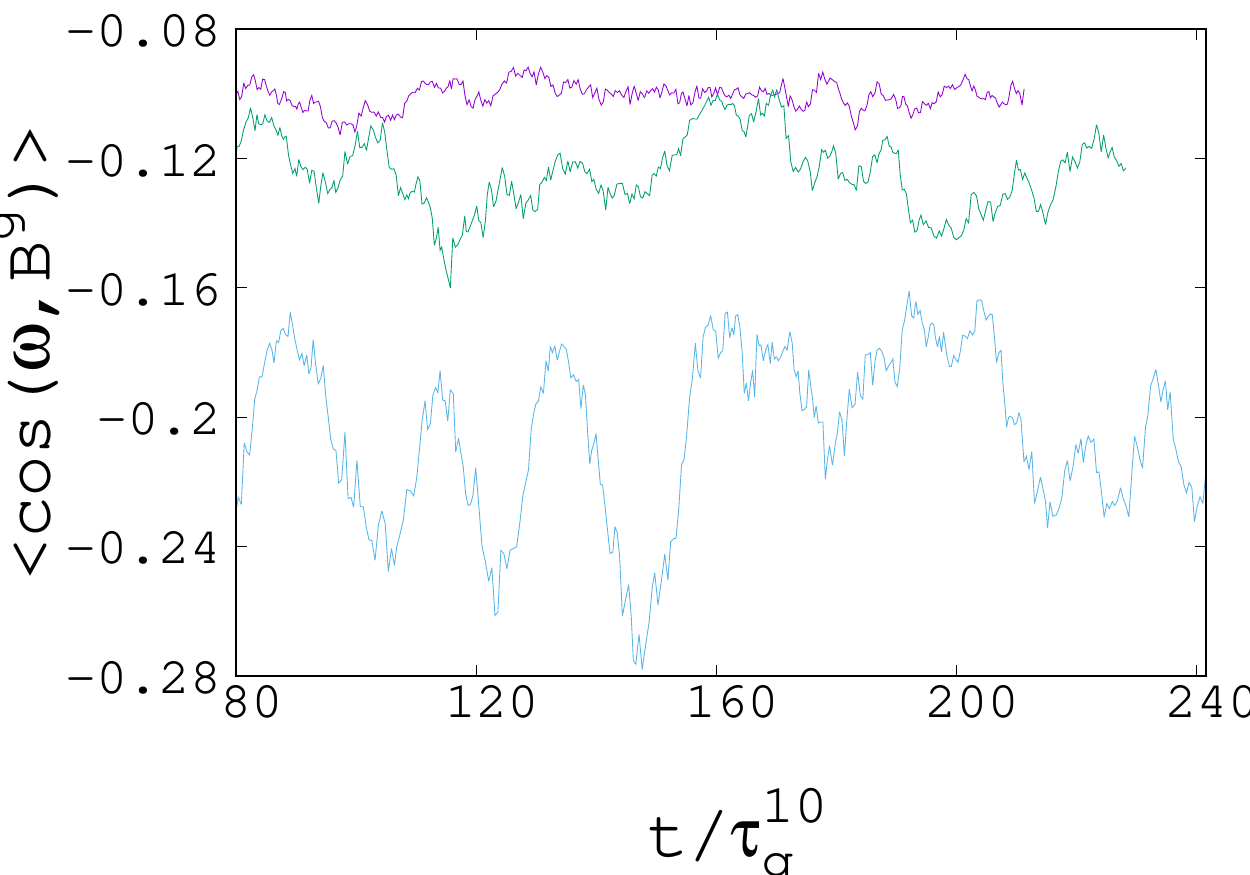}\\
\end{tabular}
\caption{\label{1}
Steady state turbulence averages for $N^g$=[10,20,40]. Left: Ratio of Lamb vector magnitude over gravitational effect magnitude. Right: Cosine of angle between $\omega$ and $B^g$.
$N^g$ increases from top to bottom curves. The shown time period corresponds to several dozens of $\tau_e^{N^g}$. 
}
\end{minipage}
\end{figure*}
All shown results correspond to steady states which are established after a transient period whose duration is {\it inversely proportional} to the coupling strength.
In full support of our scalings, the solutions inform that, as $N^g$ increases, so does the strength of gravitational effects relative
to Lamb-force effects (Fig.\ref{1}, left).
Moreover, $B^g$ and $\omega$ tend to be {\it antiparallel}, and the intensity of this effect is proportional
to $N^g$ (Fig.\ref{1}, right). This indicates that gravity tends to {\it neutralise} the Lamb force, hence, to reduce the 
vortical complexity of turbulence, and make it {\it less nonlinear}. Indeed,
the {\it highest} entstrophy and turbulence intensity values are associated with the {\it weakest} coupling (Fig.\ref{2}, left and centre). 
In addition, due to reduction of the gravitational source (vorticity) levels, strong coupling leads to smaller values for the mean square of
$\widetilde{B}^g$ (Fig.\ref{2}, right).\\

\begin{figure*}
\begin{minipage}[t]{0.99\linewidth}
\begin{tabular}[b]{ccc}
\includegraphics[width=0.33\linewidth]{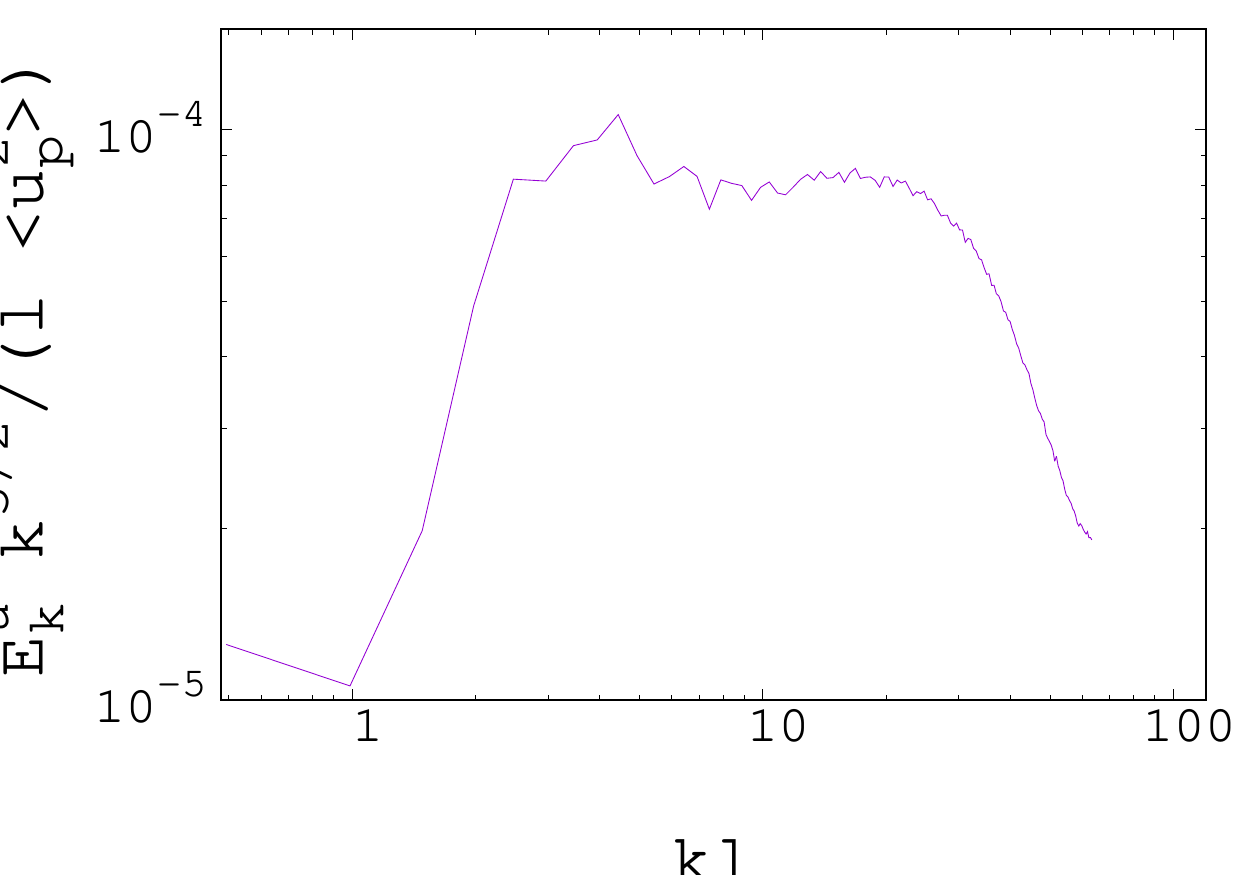}&
\includegraphics[width=0.33\linewidth]{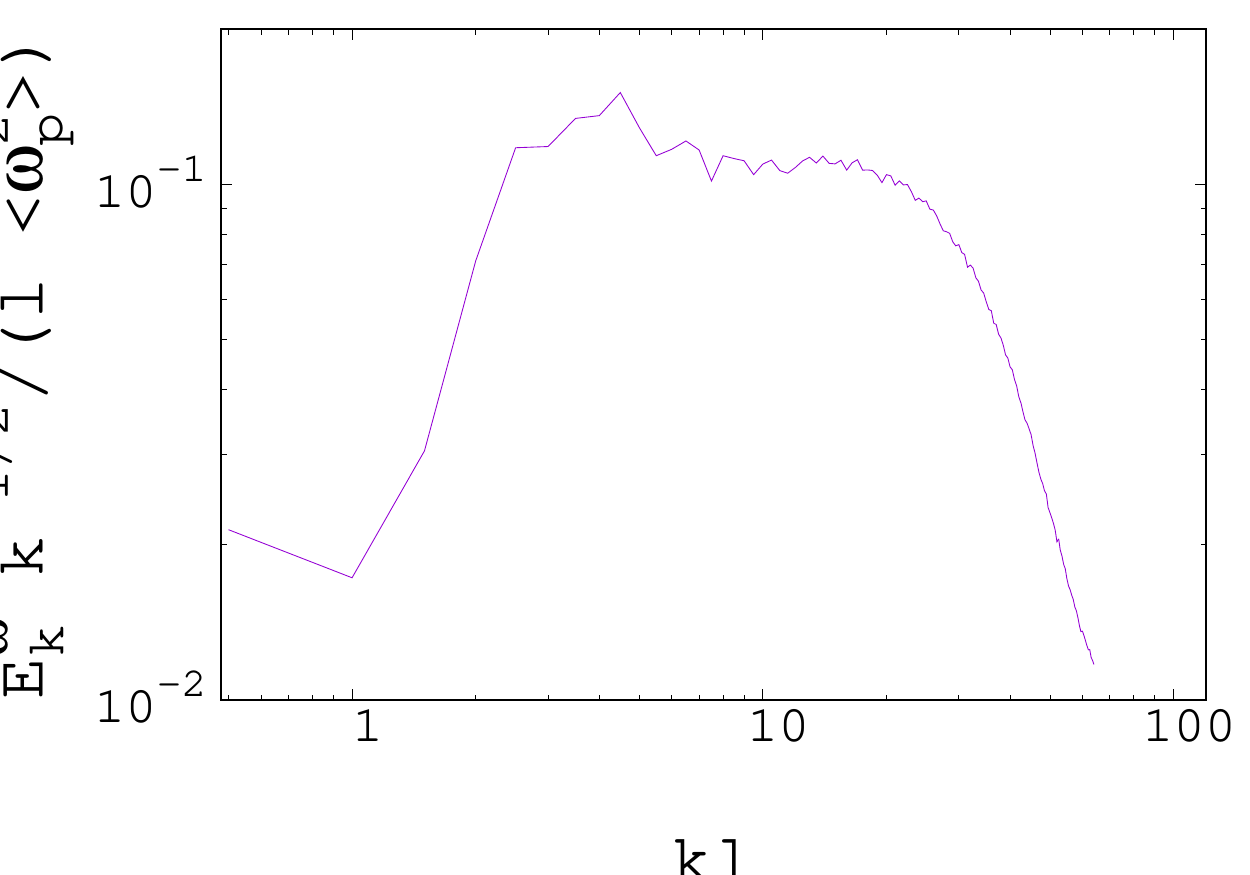}&
\includegraphics[width=0.33\linewidth]{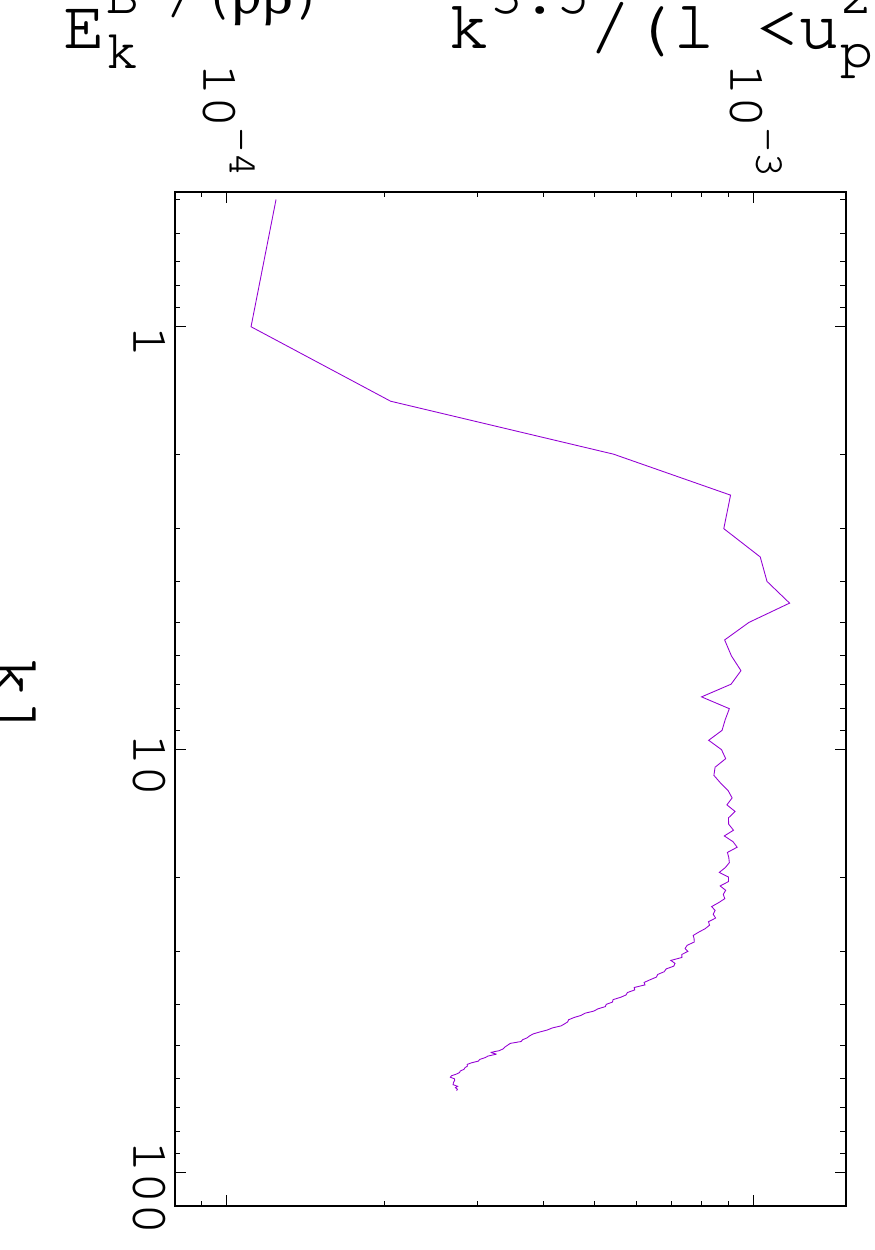}\\
\end{tabular}
\caption{\label{3}
Steady state turbulence for $N^g=20$. From left to right: velocity, vorticity and gravitomagnetic field compensated spectra,
exemplifying the corresponding $k^{-3/2}$, $k^{1/2}$ and $k^{-3.5}$ scalings. 
}
\end{minipage}
\end{figure*}

Of key importance are the velocity $E_{k}^{u}$, vorticity $E_{k}^{\omega}$, and gravitational-field $E_{k}^{\widetilde{B}^g}$ spectra.
Before enabling gravity, we verified the Kolmogorov scalings, $E_{k}^{u} \sim k^{-5/3}$ and 
$E_{k}^{\omega} \sim k^{1/3}$ for our pure turbulence calculation. Gravity induces new, GMHD scalings: $E_{k}^{u} \sim k^{-3/2}$, $E_{k}^{\omega} \sim k^{1/2}$,
and $E_k^{\widetilde{B}^g} \sim k^{-3.5}$ (Fig.\ref{3}).
The spectra shown are for $N^g=20$, since $N^g=10$ turbulence is not equally representative
of strong coupling effects, and $N^g=40$ turbulence is not equally well resolved (albeit still satisfactorily). 
The computed scalings can be understood via dimensional and exact analysis arguments.
For $E_{k}^{u}$, we can use a dimensional analytic relation due to Kraichnan \cite{biskamp} for
the rate of kinetic energy dissipation (equal to the energy-flux in wavenumber space)
$\epsilon=\tau_k (E_{k}^{u})^2 k^4$. Here, $\tau_k$ is a time scale, which is associated
with wavenumber $k$, and is characteristic of energy transfer processes from $k$ to higher wavenumbers. In Kolmogorov turbulence, $\tau_k=(E_k^M k^3)^{-1/2}$, i.e.,
the time scale of local in $k$ space turbulence eddies, which is indicative of their lifetime or the time-interval over which
fluid motions of length scale $k^{-1}$ remain correlated. However, for a self-gravitating fluid with high $N^g$,
$\tau_k$ ought to scale with $\widetilde{B}^g$, since the latter would determine the decorrelation time
via the gravitational forcing term in the NS equation. Hence, $\tau_k=(\widetilde{B}^g k)^{-1}$, and inserting this into
the expression for $\epsilon$, we obtain $E_{k}^{u}=(\epsilon \widetilde{B}^g)^{1/2} k^{-3/2}$, which agrees with the computational result.
Noting that $E_{k}^{u}$ units are $[E_{k}^{u}]=L^3 T^{-2}$, and that $[E_{k}^{\omega}]=L T^{-2}$, we obtain
$E_{k}^{\omega} \sim E_{k}^{u} k^2$, hence, $E_{k}^{\omega} \sim k^{1/2}$, which also agrees with the results.
Next, it is straightforward to predict $\widetilde{B}^g$ scaling: take the Fourier transform of $\widetilde{B}^g$ equation,
and square to obtain: $k^4 |\widehat{\widetilde{B}^g}(k)|^2=\beta \rho |\widehat{\omega}(k)|^2$. Employing the definition
$\int_{0}^{\infty} E_{k}^{\omega} dk=\iiint |\widehat{\omega}(k)|^2 d^3 k$, we deduce $|\widehat{\omega}(k)|^2 \sim k^{-3/2}$.
This gives $|\widehat{\widetilde{B}^g}(k)|^2 \sim k^{-11/2}$, and via the definition $\int_{0}^{\infty} E_{k}^{\widetilde{B}^g} dk=\iiint |\widehat{\widetilde{B}^g}(k)|^2 d^3 k$,
we obtain $E_k^{\widetilde{B}^g} \sim k^{-3.5}$, which accurately matches the computed value.\\

\begin{figure*}
\begin{minipage}[t]{0.99\linewidth}
\begin{tabular}[b]{cc}
\includegraphics[width=0.39\linewidth]{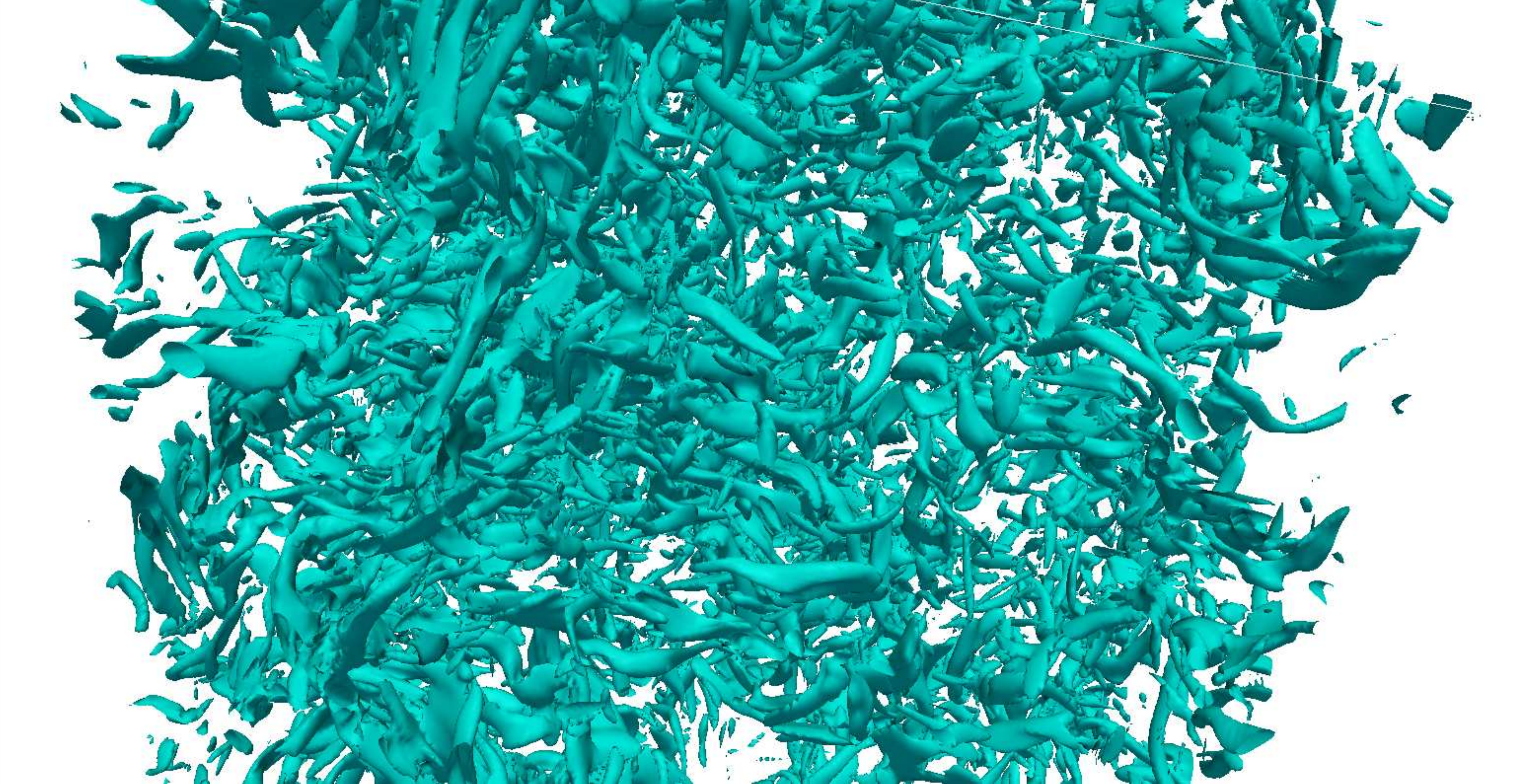}&
\includegraphics[width=0.59\linewidth]{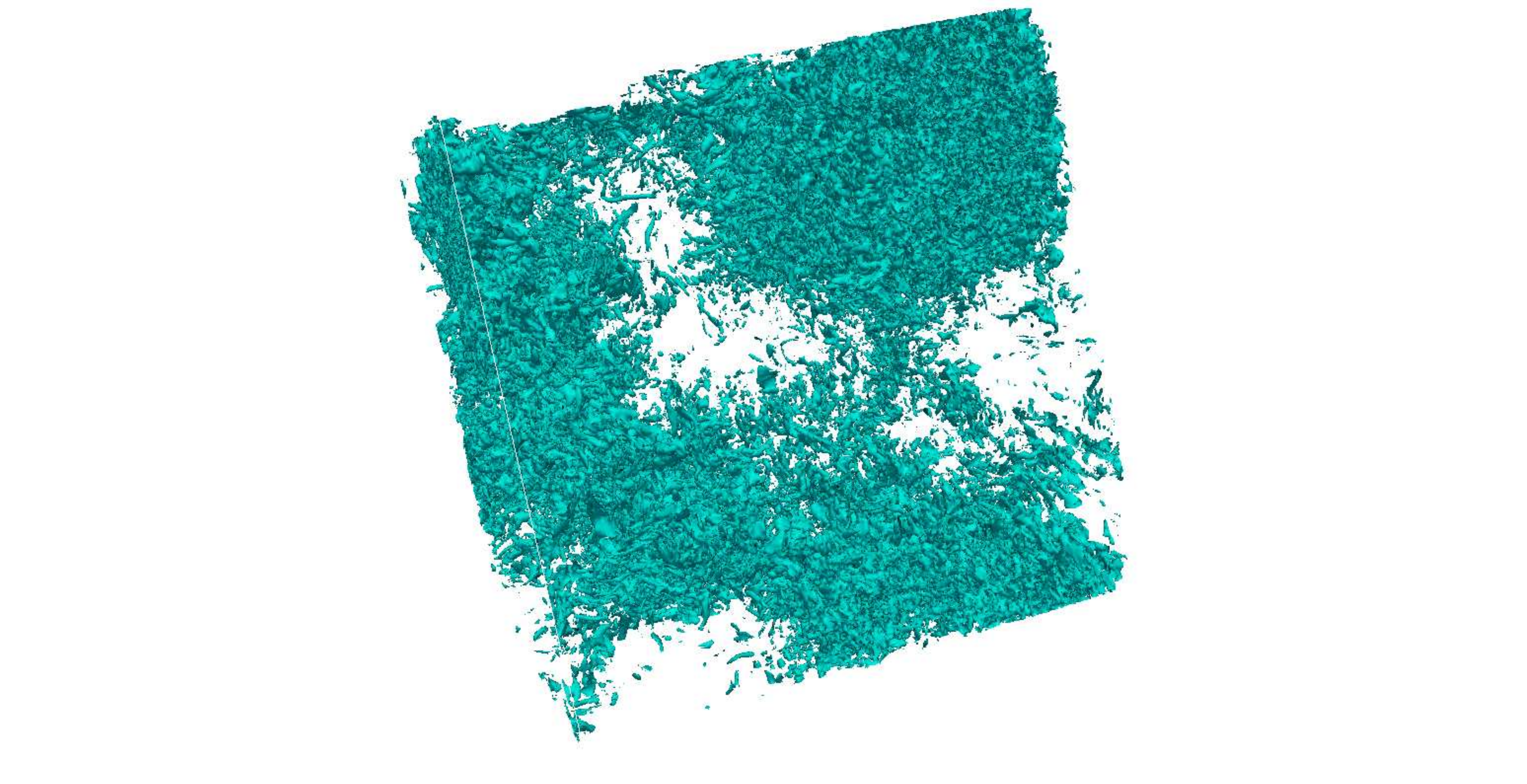}\\
\end{tabular}
\caption{\label{4}
Vorticity isosurfaces for pure turbulence (left) and gravitational turbulence for $N^g=20$ (right).
}
\end{minipage}
\end{figure*}

\begin{figure*}
\begin{minipage}[t]{0.99\linewidth}
\begin{tabular}[b]{c}
\includegraphics[width=0.59\linewidth]{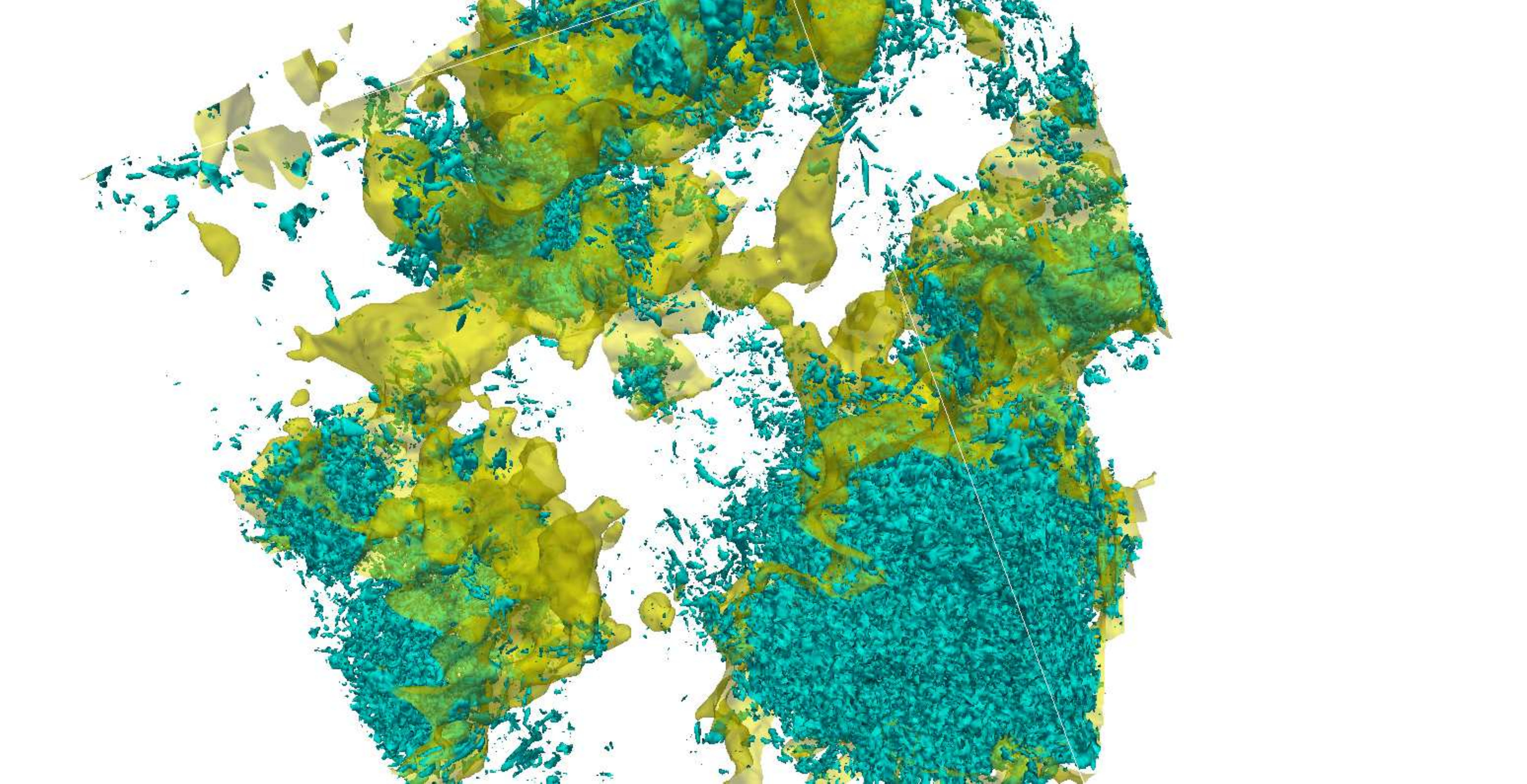}\\
\end{tabular}
\caption{\label{5}
Gravitational field (yellow) and vorticity (turquoise) isosurfaces in gravitational turbulence for $N^g=20$.
}
\end{minipage}
\end{figure*}

The morphologies of vorticity and gravitomagnetic fields present both surprising and deducible features (Figs.\ref{4}-\ref{5}). 
Pure turbulence vorticity isosurfaces drawn at $15\%$ of maximum value (Fig.\ref{4}, left)
indicate the standard, predominantly linear, NS structures, spreading homogeneously over the whole
domain. However, in the gravitational case, the vortex size is smaller, and, even when isosurfaces spanning the whole range of vorticity
levels are shown simultaneously (Fig.\ref{4}, right), large volumes devoid of any vorticity are observed. This is explained by the 
tendency of the gravitomagnetic field to neutralize the Lamb force, and suppress turbulence.
On the other hand, due to the vorticity source in the $\widetilde{B}^g$ equation, $\widetilde{B}^g$ and ${\omega}$ coexist in space,
and, since the former involves nonlocal space averages of the latter, its isosurfaces form extended structures spanning the system's domain (Fig.\ref{5}). 
Here, vorticity isosurfaces are drawn at $15\%$ of the maximum value, and magnetic field isosurfaces at $55\%$ of maximum value.
The $\widetilde{B}^g$ morphology remains the same from the smallest isolevels up to $70\%$ of maximum value, and its structures
become localized only when, at sufficiently high field values, the corresponding high-vorticity source has very small support.\\  

\section{Conclusion}

At hydrodynamic scales, relativistic gravitodynamics is characterized by three nonlinearities: (1) the nonlinearity of the Einstein equation
for the field, (2) the nonlinearity of the Navier-Stokes equation for matter, and (3) the standard nonlinearity of interacting field theory emanating
from matter-field coupling. Because of these nonlinearities, well resolved computations of relativistic self-gravitating fluids are too complex
to perform. Hence, we formulated a far simpler problem here, which, nevertheless, retains genuine relativistic effects in the guise of gravitomagnetism.
Indeed, by taking the weak-field, slow-motion limit, we removed the first nonlinearity, but preserved the other two. 
Although this appears to be a limitation, we still have a very demanding mathematical problem in our hands, 
especially when the fluid is turbulent. This
is because turbulence is a unique example of nonlinearity that can become arbitrarily strong without ``breaking"
the underlying system (a fluid can sustain extremely large Reynolds numbers). Moreover, turbulence physics are dominated by 
Biot-Savart interactions between vortical structures, i.e., by vorticity, which also is the source of the gravitomagnetic field. 
In other words, the sources of gravity are the very structures that dominate turbulence physics. It was shown that the third nonlinearity
tames the second. Indeed, enstrophy is intensified by vortex stretching and peaks at high wavenumbers, as a result of Lamb-force driven turbulence
kinetic energy cascade. But as the cascade intensifies small-scale enstrophy, the latter generates a gravitomagnetic field whose action on the fluid
counterbalances the Lamb force driving the cascade. In other words, gravitomagnetism tends to linearize and damp out turbulence.
At statistical equilibrium, the field levels are consistent with enstrophy intensification in turbulence that is allowed by
the degree of flow nonlinearity reduction due to these field levels. The latter are inversely proportional to the field-matter coupling.\\

Future elaboration of GMHD vortex dynamics and detailed probing of gravity mediated
vortex interactions in turbulence could be informative. Gravitomagnetic effects would generate novel 
coherent-structure formation mechanisms, and alter strain-rate tensor related statistics. 
Finally, the employment of advanced geometrical \cite{louise} and topological \cite{topos,pawel} methods for the
characterization of gravitational and vorticity field structures would help indicate in a quantitative (rather than visual)
way the differences between pure and gravitomagnetic turbulence flow patterns.\\

\bibliographystyle{plain}

\end{document}